\documentclass{ws-ijmpe}
\usepackage[super,compress]{cite}
\usepackage{color}
\usepackage{graphicx}
\begin{document}

\markboth{Riou Nakamura et al.}
{Big-Bang nucleosynthyesis}

\catchline{}{}{}{}{}

\title{{Big-Bang nucleosynthyesis: constraints on nuclear reaction rates,
 neutrino degeneracy,
inhomogeneous and Brans-Dicke models}
}

\author{\footnotesize RIOU NAKAMURA
}
\address{
Kurume Institute of Technology, Kamitsu-machi, Kurume, 830-0052,
Japan, \\
Department of Physics, Kyushu University, Motooka,
Fukuoka 819-0395, Japan
\\
riou@astrog.phys.kyushu-u.ac.jp}

\author{MASA-AKI HASHIMOTO}
\address{Department of Physics, Kyushu University, Motooka,
Fukuoka 819-0395, Japan
}

\author{RYOTARO ICHIMASA}
\address{Department of Physics, Kyushu University, Motooka,
Fukuoka 819-0395, Japan
}

\author{KENZO ARAI}

\address{Department of Physics, Kumamoto University, Kurokami,
Kumamoto 860-8555, Japan
}

\maketitle

\begin{history}
\received{Day Month Year}
\revised{Day Month Year}
\end{history}

\begin{abstract}
We review the recent progress in the Big-Bang nucleosynthesis which includes
the standard and non-standard theory of cosmology, effects of neutrino degeneracy, and 
inhomogeneous
nucleosynthesis  within the framework of a Friedmann model. As for a non-standard
theory of gravitation, we adopt a Brans-Dicke theory which incorporate
a cosmological constant.   
We constrain various parameters associated with each subject。
\end{abstract}                                

\ccode{PACS Nos.:98.80.-k; 26.35.+c; 98.80.Ft}

\section{Introduction}
\label{sec:ibbn_intro}

Big-Bang nucleosynthesis (BBN) is one of the best astrophysical sites to understand the origin
of the light elements,\cite{alpher1948} because it produces the primordial
abundances of $^4$He, D and $^7$Li \cite{Luridiana2003,perm2007}.
 The simplest and most trustful theory is called the standard Big-Bang
 nucleosynthesis (SBBN), where we assume the cosmological principle that
 the universe is homogeneous and isotropic in a large scale\cite{Iocco:2008va}.
Unfortunately, current situation of SBBN is incompatible with the observed abundance of
$^7$Li, 
although it has been pointed out that observational determinations of the
${}^7$Li abundance contain uncertainties associated with an adopted
atmospheric model of
metal-poor stars\cite{Ryan2000}  
or some unknown physical processes\cite{Coc2004}.

On the other hand, heavy element nucleosynthesis
beyond the mass number $A=8$ has been investigated
in a framework of inhomogeneous BBN (IBBN)
~\cite{TerasawaSato89,IBBN0,IBBN1,Alcock1987,Jedamzik1994,Matsuura:2004ss}.
The baryon inhomogeneity could be originated from baryogenesis\cite{Matsuura:2004ss} or some phase
transitions~\cite{Alcock1987,Fuller1988} which occur as the universe cools down 
during the expansion. 
It should be noted that IBBN motivated by QCD phase transition encounters difficulty, 
because the transition has been
proved to be crossed over  by the Lattice QCD simulations\cite{aoki}, 
which means that the phase transition occurs smoothly between the
quark-gluon plasma and hadron phase under the finite temperature and zero chemical potential.
Although a large scale inhomogeneity of baryon distribution should be ruled out by cosmic microwave
background (CMB) observations~\cite{WMAP3,WMAP5}, 
small scale inhomogeneities are still allowed
 within the present accuracy of observations. 
Therefore, it is possible 
for IBBN to occur in some degree during the early era.

Considering recent progress in observations for peculiar abundance
distributions, it is worthwhile to re-investigate
IBBN\cite{Moriya2010,Lara2006,Vonlanthen2009}
It has been found that\cite{Matsuura2005} 
synthesis of heavy elements
for both $p$- and $r$-processes is possible if $\eta > 10^{-4}$
and that the high
$\eta$ regions are compatible with the observations of the light elements,
$^4$He and D.
However, the analysis is only limited to a parameter of a specific
baryon number 
concentration, and a wider possible parameter region should be
considered.
Therefore we adopt a two-zone model and derive constraints comparing the
BBN result with available observations.

In the meanwhile, whole validity of general relativity has not been
proved. For example, some kinds of scalar fields may exist in the
present epoch. Therefore, 
we need to examine whether a possibility for a non-standard theory of
gravitation survives from the viewpoint of BBN.
Simple version of a scalar field theory is the Brans-Dicke theory
which include a scalar field. This theory leads to a variation of
the gravitational constant and induces an acceleration of the universe.
Therefore, we show compatibility between the observations of light
elements and the Brans-Dicke theory with a $\Lambda$ term.

In section 2, we give the framework of SBBN and present a detailed 
comparison among observations and Big-Bang nucleosynthesis in section 3. 
As an approach to explain a wide range of observational He abundances,
we give the effects of
neutrino degeneracy on the production of He in section 4. 
Section 5 is devoted to
nucleosynthesis with use of a simple approach for the inhomogeneous BBN. Finally,
we give non-standard BBN based on the Brans-Dicke theory in section 6 which remains a
room beyond the general relativity. Section 7 summaries possible
variations of BBN.

\section{Thermal evolution of the universe}
In this section, we summarize the evolution of the standard cosmological
model, i.e. Friedmann model based on the cosmological principle or the
Friedmann-Robertson-Walker metric\footnote{We adopt the system in units of $c=1$.},
\[
 ds^2_{}=g^{}_{\mu\nu}dx^{\mu}_{}dx^{\nu}_{}=-dt^2_{}+a^2(t)\left(
 \frac{dr^2}{1-kr^2}+
r^2d\theta^2_{}+r^2\sin^2{\theta}d\phi^2\right), 
\]
where $a(t), t$, and $k$ are the cosmic scale factor, the cosmic time, and the
spatial curvature, respectively.
The Friedmann model 
has been constructed with use of the Einstein equation,
\[
R^{}_{\mu\nu}-\frac{1}{2}g^{}_{\mu\nu}R=8\pi GT^{}_{\mu\nu}. 
\]
Here $R^{}_{\mu\nu}, R, G$ and $T^{}_{\mu\nu}$ are the Ricci tensor, the
scalar curvature, the gravitational constant, and the energy-momentum
tensor of the perfect fluid written as $T_{\mu\nu}=pg_{\mu\nu}+(p+\rho)u_{\mu}u_\nu$, 
where $p$ is the pressure, $\rho$ is the energy density, and
$u_\mu$ is the four velocity.

In practice, we can follow the evolution of temperature $T$ and energy density $\rho$ by 
solving the Friedmann equation
\begin{equation}
H^2\equiv\left(\frac{\dot{a}}{a}\right)^2 = 
 \frac{8\pi G}{3}\rho -\frac{k}{a^2_{}},
\label{eq:friedman_equation}
\end{equation}
where $H$ is called as the Hubble parameter.
The total energy density $\rho$ is written as 
\[
\rho=\rho^{}_{\gamma}+\rho^{}_{\nu}+\rho^{}_{e^{\pm}}+\rho^{}_{b} +
\rho^{}_{\rm DM} + \rho^{}_{\Lambda},
\]
where the subscripts $\gamma,~\nu,~e^{\pm},~b,~{\rm DM}$ and $\Lambda$ indicate 
photons, neutrinos, electrons and positrons, baryons, dark matter, and the
vacuum energy, respectively.

The energy conservation law, ${T^{\mu\nu}_{}}_{;\mu}=0$, reduces to 
\begin{equation}
\frac{d}{dt}(\rho a^3) + 
p\frac{d}{dt}(a^3)
=0.
\label{eq:rho_evolution}
\end{equation}
The equation of state is described as 
\[
 {p}/{\rho} = 
 \begin{cases}
  1/3 & \text{for photons and neutrinos,} \\
  0 & \text{for baryons and dark matter,} \\
 -1 & \text{for vacuum energy.}
 \end{cases}
\]
Therefore, we obtain the evolution of the energy density as 
$\rho_{b}$ and $\rho^{}_{\rm DM}\propto a^{-3}_{}$, and 
$\rho^{}_{\gamma, \nu}\propto a^{-4}$.
The temperature varies as $T^{}_{\gamma}\propto a^{-1}$ 
except for the era of the significant entropy transfer from $e^{\pm}_{}$
to photons 
at $T\simeq 5\times10^9$~K. 

Usually, the present values of the energy densities are expressed as the
density parameter $\Omega^{}_{i}=\rho^{}_{i,0}/\rho^{}_{cr}$ for
$i=\gamma, \nu, b$, DM, and $\Lambda$ with the critical density
$\rho^{}_{cr}=3H^2_0/8\pi G$. Here the subscript $0$ means the value at
the present time. 

For the baryon density, we also use the baryon-to-photon ratio $\eta$,
\begin{equation}
\eta^{}_{}\equiv \frac{n^{}_{b}}{n^{}_{\gamma}} = 2.74\times 10^{-8}_{}\Omega^{}_{b}h^2,
\end{equation}
where $h$ is the dimensionless Hubble constant:  $h = H^{}_{0}/100$~[km/s/Mpc].
The value of $\eta$ is kept constant after the electron-positron pair annihilation, because
the number densities of photons $n^{}_\gamma$ and baryon $n^{}_b$ vary as
$n^{}_{\gamma}$ and $n^{}_{b}\propto a^{-3}$. 
From the\ results of observation of CMB by
the Planck satellite, $0.02180<\Omega^{}_bh^2 < 0.02272$ at $95$\% 
confidence levels~(C.L.)~\cite{Planck2015cosmo},
we obtain the range  
\begin{equation}
 5.96< \eta^{}_{10} < 6.22 \label{eq:eta_planck}
\end{equation}
where $\eta^{}_{10}= \eta/10^{-10}$.

\section{Standard Big-Bang nucleosynthesis}
\subsection{Physical process of the Big-Bang nucleosynthesis}

In this section, we describe the standard model of BBN. 
At the stage $T > 10^{10}$~K, photons, neutrinos, and electron (plus positron) are 
dominant. 
The energy density is written as follows:
\[
\rho = 
\frac{\pi^2k^{4}_{\rm B}}{30\hbar^3}\left( 2+ \frac{7}{2}+\frac{7}{4}N_\nu \right)T^4, 
\]
where $k^{}_{\rm B}$, $\hbar$, and $N_{\nu}$ are the Boltzmann constant,
 the Planck constant reduced by $2\pi$, and the effective number  of
 neutrinos (we fixed the value $N_\nu = 3$). 
The first, second, and third 
terms are contributions 
of photons, electrons
and positrons, 
and neutrinos,
respectively.
 From Eq.\eqref{eq:friedman_equation}, the expansion rate of the Universe is 
$H\sim G^{1/2}T^2$.
At that time, neutrons and protons are coupled through the weak interaction
as follows:
\begin{eqnarray}
n + e^+ & \longleftrightarrow & p + \bar{\nu}_e, \notag \\
n + \nu_e & \longleftrightarrow & p + e^-,   \label{eq:weak_reac} \\
n & \longleftrightarrow & p + e^- + \bar{\nu}_e \notag
\end{eqnarray}
The rate of the weak interaction $\Gamma_{wk}$ can be written as 
$\Gamma_{wk}\sim G_FT^5$. Here $G_F$ is the Fermi coupling constant.
The number ratio of the neutrons to protons 
is written as 
\begin{equation}
\frac{n^{}_n}{n^{}_p}=\exp{\left( -\Delta m/k_{\rm B}T\right) }, 
\label{eq:npratio}
\end{equation}
where $\Delta m$ is the mass difference between the neutron and proton:
$\Delta m=m_n-m_p = 1.29$~MeV.

When $\Gamma^{}_{wk}\simeq H$ at $k_{\rm B}T\simeq 1$~MeV, the ratio \eqref{eq:npratio} 
is frozen (except for neutron's $\beta$-decay) and the ratio of the neutrons and protons
is fixed to be $1/7$. The ratio in this epoch is very important.
As shown later, 
the most abundant nuclide except for ${}^1_{}$H in the BBN epoch 
is ${}^4$He. 
The abundance of ${}^4$He is defined by the ratio as follows: 
\[
Y_{\rm p} = \frac{\rho^{}_{{}^4 \rm He}}{\rho^{}_b} \sim \frac{2n^{}_n}{n^{}_p + n^{}_n}\sim 0.25
\]
where $Y_{\rm p}$ is the mass fraction of ${}^4_{}$He.

\begin{figure}[tb]
\begin{center}
 \includegraphics[width=0.5\linewidth,keepaspectratio,clip]{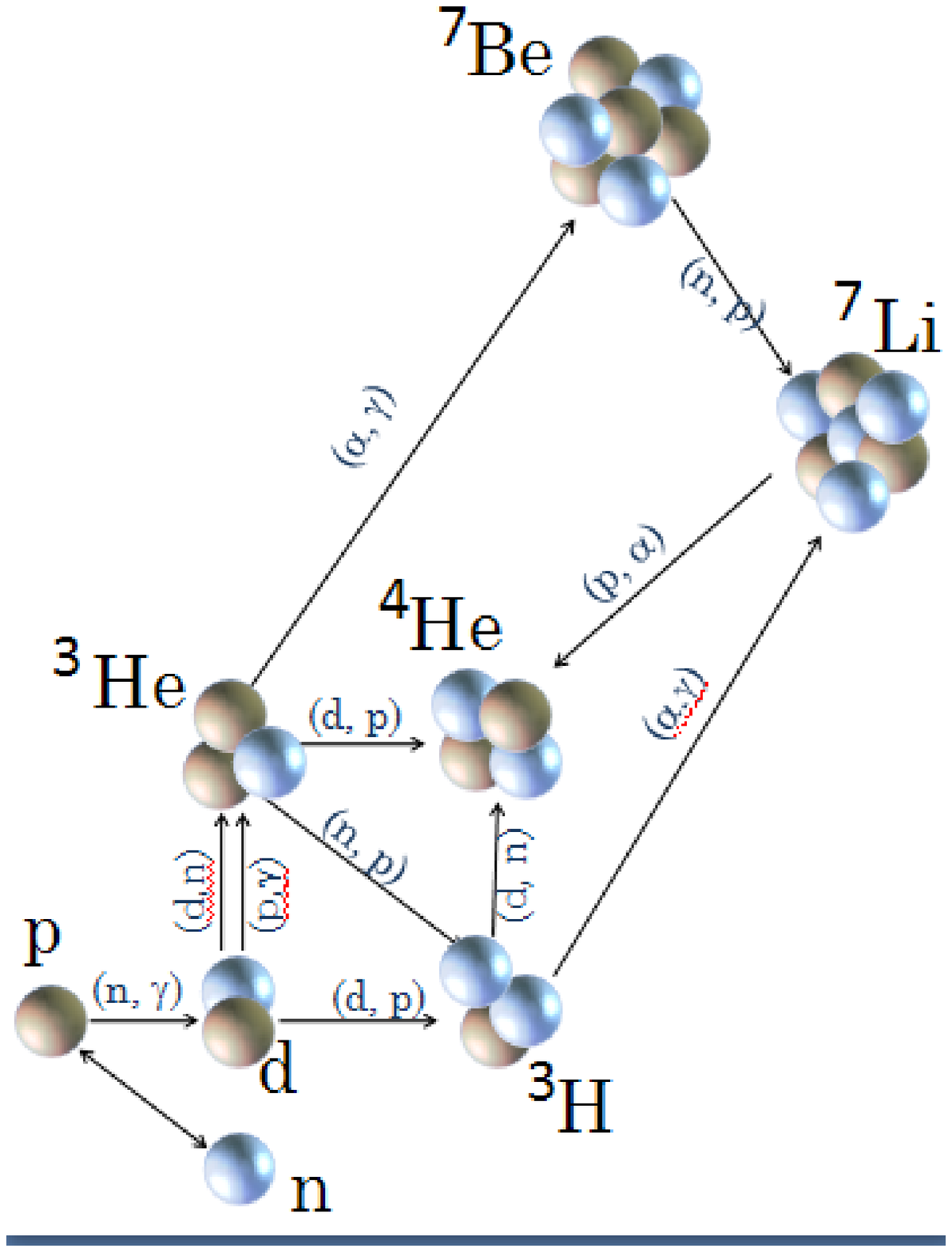}
\end{center}
\vspace{-0.3cm}
\caption{Nuclear reaction network in Big-bang nucleosynthesis.}
\label{fig:bbnewkl}
\end{figure}

\begin{figure}[tb]
\begin{center}
 \includegraphics[width=0.9\linewidth,keepaspectratio]{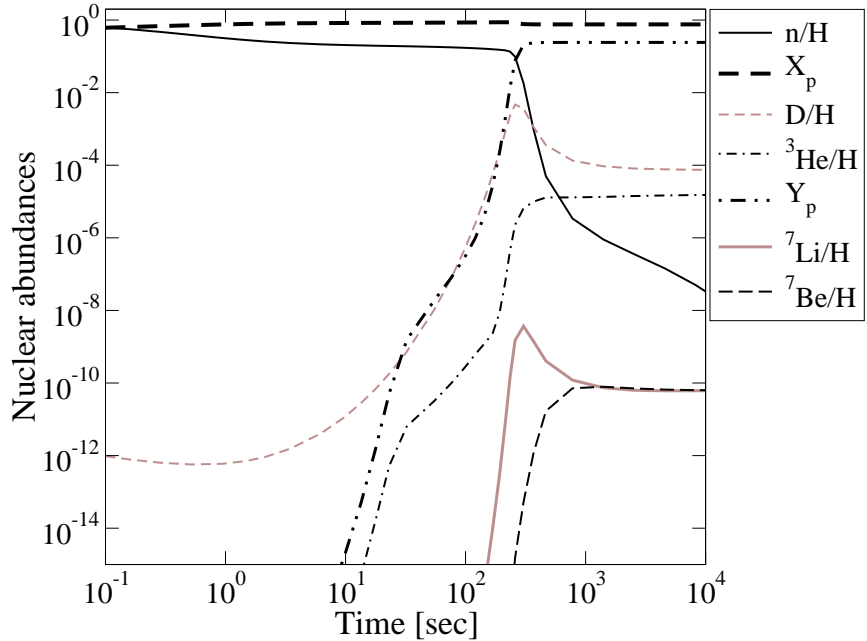}
\end{center}
\vspace{-0.3cm}
\caption{
Evolution of mass fractions of ${}^1_{}$H ($X_{\rm p}$) and ${}^4_{}$He
 ($Y_{\rm p}$) and number abundances of other nuclides in SBBN.
}
\label{fig:bbnevol}
\end{figure}

When the temperature reaches to $0.1$ MeV, 
the synthesis of light elements starts
by the neutron capture reaction of protons: 
n + p $\longrightarrow$ D $+\gamma$. 
Figure \ref{fig:bbnewkl} illustrates the 12 important reactions of BBN.

\begin{figure}[tb]
\begin{center}
\includegraphics[width=0.7\linewidth,keepaspectratio]{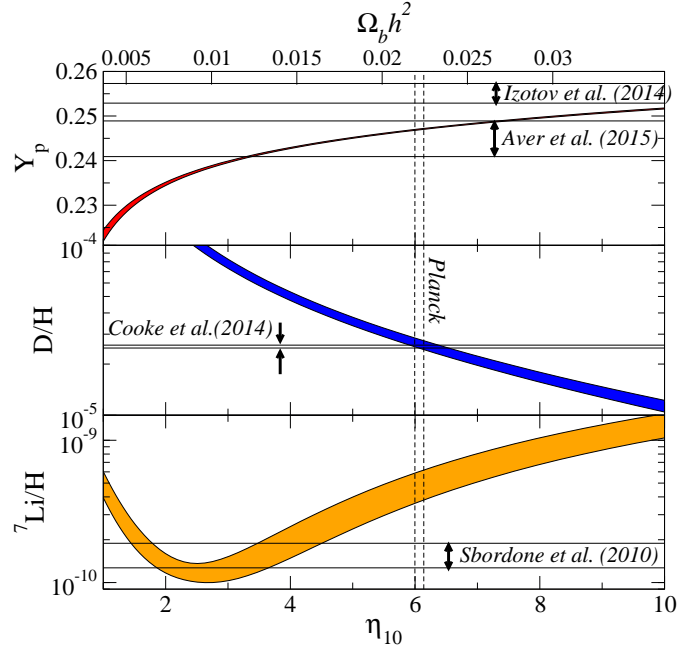}
\end{center}
\caption{Abundance of light elements produced in SBBN as a function of $\eta$.
The vertical region $\Omega^{}_bh^2=0.02222 \pm 0.00023$ indicates the constraint from Planck\cite{Planck2015cosmo}. 
}
\label{fig:sbbneta}
\end{figure}

Figure \ref{fig:bbnevol} shows the numerical result of BBN  
with $\eta^{}_{10}=6.19$ and the neutron life time $\tau^{}_{\rm n} = 880.1$ sec\cite{PDG2012}.
In the BBN era, the heavy element such as CNO cannot be synthesized,
because there are no stable element of the mass number $A=5$ and $A=8$.  
In Fig.~\ref{fig:sbbneta} we show the produced final values of $Y^{}_{\rm p}$, D/H and $^7$Li/H
as a function of $\eta$. The line widths for individual elements correspond to
 the errors attached to the nuclear reaction rates of NACRE II\cite{NACRE2} .
Although a new decay rate of free neutrons 
 $\tau^{}_{\rm n} = 880.1 \pm 1.1$ sec yields\cite{PDG2012} $Y_{\rm p}\simeq 0.245$, we adopt the conservative rate considering the uncertainty of the half life.\cite{sereb}.
When the baryon density is high, more D can be synthesized from neutrons and protons.
However, reactions which yield helium also begin. As a result, ${}^4$He increases while D decreases as shown in Fig.~\ref{fig:sbbneta}. 
We note that 
the ${}^7_{}$Li abundance decreases with increasing $\eta$
for $\eta^{}_{10} < 3.0$. For $\eta^{}_{10} > 3.0$,  the tendency
becomes opposite. For high baryon density, ${}^7$Be can be 
synthesized through ${}^4$He capture reaction on ${}^3$He, and
consequently, it 
is converted to ${}^7_{}$Li through the electron capture reaction after
recombination of ${}^7_{}$Be.

\subsection{Nuclear reaction rates}
To obtain primordial abundances of the light elements numerically shown
 in Fig.\ref{fig:bbnevol}, 
 it is necessary to solve differential equation of two body reaction for a number fraction $y_k$ as follows:
\[
\frac{dy_k}{dt}= \sum_{i,j,l}{y_iy_j [ijkl]} - \sum^{}_{i,j,l}{y_ky_l[klij]}, 
\]
where $[ijkl]\def \rhoN_A<\sigma v>_{ij}$ means the rates of a
thermonuclear reaction: $i + j \rightarrow k + l$. Usually the rate is a
function of the baryon density and temperature.
The other main channel responsible for BBN is the transformation between
n and p. We note that three body reactions can be negligible in SBBN.

For BBN study in these 10 years, the reaction rates 
given by Descouvemont et al.\cite{DAA04} (DAA) and 
NACRE II.\cite{NACRE2} have been used. 
The DAA rates are obtained by adopting the $R$-matrix theory so as to 
fit with
the low energy data. The NACRE II rates are the updated version of the
NACRE compilation\cite{NACRE1999} which include the thermonuclear
reaction rates obtained experimentally for nuclei with A $\le$ 16.
We compare the BBN results between DAA and NACRE II.

\begin{figure}[tb]
\begin{center}
\includegraphics[width=0.7\linewidth,keepaspectratio]{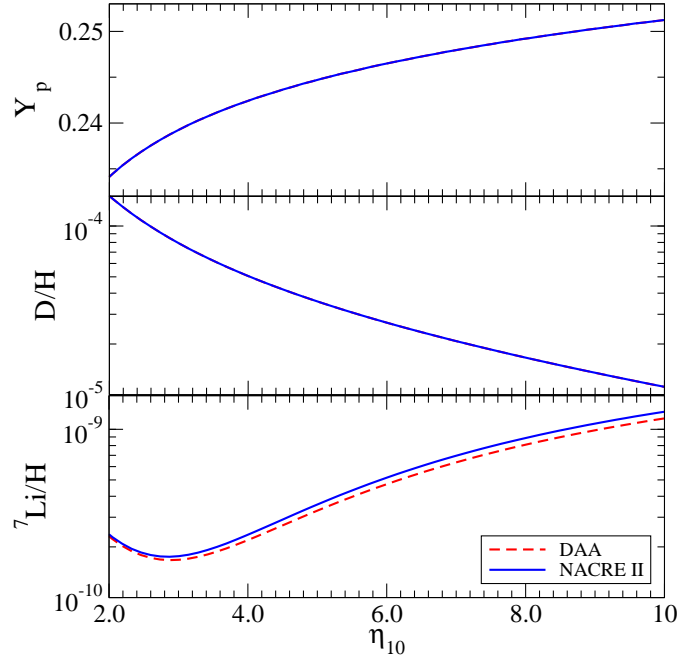}
\end{center}
\caption{Produced abundances of light elements. The solid curves are calculated from the rates of NACRE II and the dashed ones are of DAA.
}
\label{fig:nacre2vsdaa}
\end{figure}

\begin{figure}[tb]
\begin{center}
\includegraphics[width=0.7\linewidth,keepaspectratio]{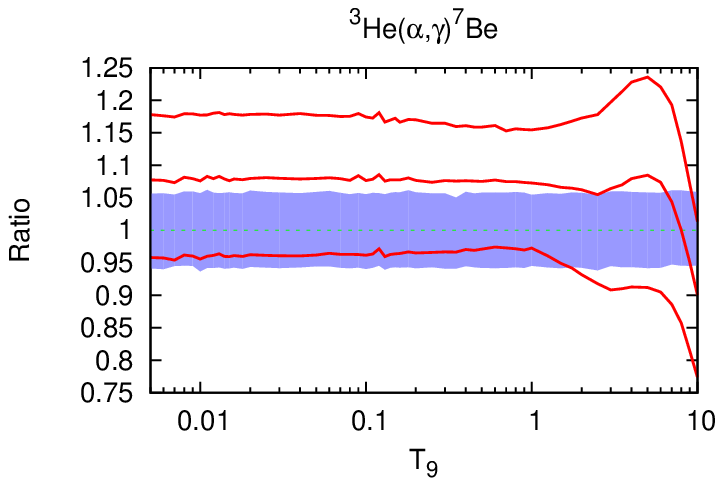}
\end{center}
\vspace{-0.5cm}
\caption{Rates of ${}^3$He ($\alpha, \gamma$)${}^7$Be from NACRE II relative to
 DAA (adopted value). The horizontal axis
means the temperature normalized by $10^{9}_{}$~K.  The shaded area
 indicates the uncertainty of the reaction rate by DAA.
The red curves correspond to 
the high (upper curve) adopted (middle curve) and low (lower curve)
 values of reaction rates in NACRE II, respectively.
 }
\label{fig:be7_nacre2vsdaa}
\end{figure}

Figure~\ref{fig:nacre2vsdaa} shows the results of
BBN calculated by using the reaction rates of DAA and NACRE II. The
 abundances of ${}^4$He and D are almost the same.
However, ${}^7$Li by NACRE II is 0.5 \% higher than that by DAA. This is
because, as shown in Fig.\ref{fig:be7_nacre2vsdaa}, the rate of the reaction ${}^3$He$(\alpha, \gamma){}^7$Be of NACRE II is higher than that of DAA at the temperature range of BBN.

\subsection{Constraints from observations of ${}^{4}$He, D, and ${}^7Li$}

The observation of ${}^4$He is important to constrain  physics of the early universe, 
because the theoretical value of ${}^4$He is sensitive to cosmological models, 
such as the gravitational model beyond general relativity and effective neutrino number.

Helium-4 is the most abundant element 
nuclide except for ${}^1_{}$H produced in BBN.
It is also yielded through hydrogen burning inside stars. Since the
abundance of ${}^4$He grows after BBN era, its observed value is an
upper limit to the primordial abundance.
Thence,  it is not easily to determine primordial  $Y_{\rm p}$. 
However, it is deduced that primordial ${}^4$He remains in a
low-metallicity region, because star formation does not occur there.　
Then ${}^4$He is determined from the recombination lines  of proton and
helium in extragalactic HII regions.

Recently, observational abundance of ${}^4$He has a conflict between two groups.
Izotov et al. report 
\begin{equation}
Y_{\rm p}=0.254 \pm 0.003 \label{eq:HeIzotov2013}
\end{equation}
using linear relation $Y^{}_{}-$O/H for 111 highest-excitation HII regions\cite{Izotov2013}.
And adding the HeI $\lambda$ 10830 \AA ~ emission line, they derived\cite{Izotov2014}
\begin{equation}
Y_{\rm p}=0.2551 \pm 0.0022. \label{eq:HeIzotov2015}
\end{equation}
On the other hand, 
using A Markov Chain Monte Carlo analysis for 93 HII regions, 
Aver et al.\cite{Aver2013} reported 
\begin{equation}
Y_{\rm p}=0.2465 \pm 0.0097. \label{eq:HeobsAver2013}
\end{equation}
Also they reported\cite{Aver2015}
\begin{equation}
Y_{\rm p}=0.2449 \pm 0.0040. \label{eq:HeobsAver2015}
\end{equation}
by evaluating the effects of adding He I $\lambda$10830 infrared emission line 
in helium abundance determination.

{Helium-3 is made from deuteron 
by D(p,$\gamma$) ${}^3$He in the star.
Therefore, observations of D give a upper limit of primeval values.
Since the abundance of D strongly depends on the baryon density, 
D is called as ``baryometer".}
The abundance of D is 
measured in 
QSO absorption line system at high-redshift.
Recently, the primordial D abundance is determined with a high-accuracy.
Pettini \& Cooke reported\cite{Pettini2012}
\begin{equation}
\text{D/H}  =(2.535 \pm 0.05)\times 10^{-5}. \label{eq:DobsPettini2012}
\end{equation}
from the metal-poor damped Lyman $\alpha$ (DLA) system at $z=3.05$.
And Cooke et al. reported\cite{Cooke2014}
\begin{equation}
\text{D/H}  =(2.53 \pm 0.04)\times 10^{-5}. \label{eq:DobsCooke2014} 
\end{equation}
from the very metal-poor DLA toward SDSS J1358+6522.


{{
The primeval abundance of ${}^7$Li can be estimated from observations of Population  II stars in our galaxy. }
In the old galaxy, the relation between [Li/H]  and [Fe/H] becomes constant at a low 
metallicity star~(it is called  ``Spite-Plateau").  
The observed abundance of $^7$Li in Population II
stars is given by 
Sbordone et al. ~\cite{sbord}:
}
\begin{equation}
 {\rm {}^{7}Li/H} = \left( 1.58 \pm 0.31 \right) \times 10^{-10}.
\label{eq:Li7_Sbord} 
\end{equation}

To find reasonable values of $\eta^{}_{10}$ which satisfy the consistency between
BBN and observed values, we calculate $\chi^{2}_{}$ as follows:
\begin{equation}
 \chi^{2}_{i}(\eta)={\frac{\left( Y^{th}_i(\eta) -
			     Y^{obs}_{i}\right)^2}
{{\sigma^{2}_{th,i}}+{\sigma^{2}_{obs,i}}}}, 
\label{eq:chisq}
\end{equation}
where $Y^{}_{i}$ and $\sigma^{}_i$ are the abundances and their uncertainties
for elements $i~(i=Y_p, {\rm D},{}^7{\rm Li})$, respectively.
The value $\sigma_{th,i}$ is obtained from the Monte-Carlo calculations using 1$\sigma$ errors associated with 
nuclear reaction rates. 
The observational values, $Y^{obs}_i$ and their errors $\sigma_{obs,i}$, are taken 
from \eqref{eq:HeIzotov2015}, \eqref{eq:HeobsAver2015}, \eqref{eq:DobsCooke2014}, and \eqref{eq:Li7_Sbord}.

Finally, we obtain the range of $\eta_{10}$ with $1\sigma$ C.L. from individual observations:
\begin{eqnarray}
&3.28 \le \eta_{10} \le 7.27 & \hspace{0.5cm} \text{from Aver et al.\cite{Aver2015}}, \notag  \\
&10.1 \le \eta_{10} \le 18.2 & \hspace{0.5cm} \text{from Izotov et al.\cite{Izotov2014}},   \notag\\
&6.08 \le \eta_{10} \le 6.38 & \hspace{0.5cm} \text{from Cooke et al.\cite{Cooke2014}}, \notag \\
&3.28 \le \eta_{10} \le 4.40 & \hspace{0.5cm} \text{from Sbordone et al.\cite{sbord}}  \notag
\end{eqnarray}
As against the excellent agreement for D and $^4$He, the discrepancy for
${}^7$Li is unallowable. This is called  "${}^7$Li problem". Since the
baryon density is determined
precisely from CMB observation by WMAP and/or Planck, this problem is conspicuous
more than ever. 

To solve this problem, 
many possibilities have been considered,
such as unstable massive particles\cite{Kusakabe2011, KohriTakayama2006}, 
nuclear reaction rates coupled with other reaction paths\cite{Broggini2012,Coc2012}, 
and  a scalar-tensor theory of gravity~\cite{Coc2006rt,Larena2007}.
It is noted that the observations of ${}^7$Li involve uncertain
atmospheric models concerning the low metallicity stars\cite{Korn2006} . 

 
\section{Neutrino degeneracy}
\begin{figure}[tb]
\resizebox{\hsize}{!}{\includegraphics{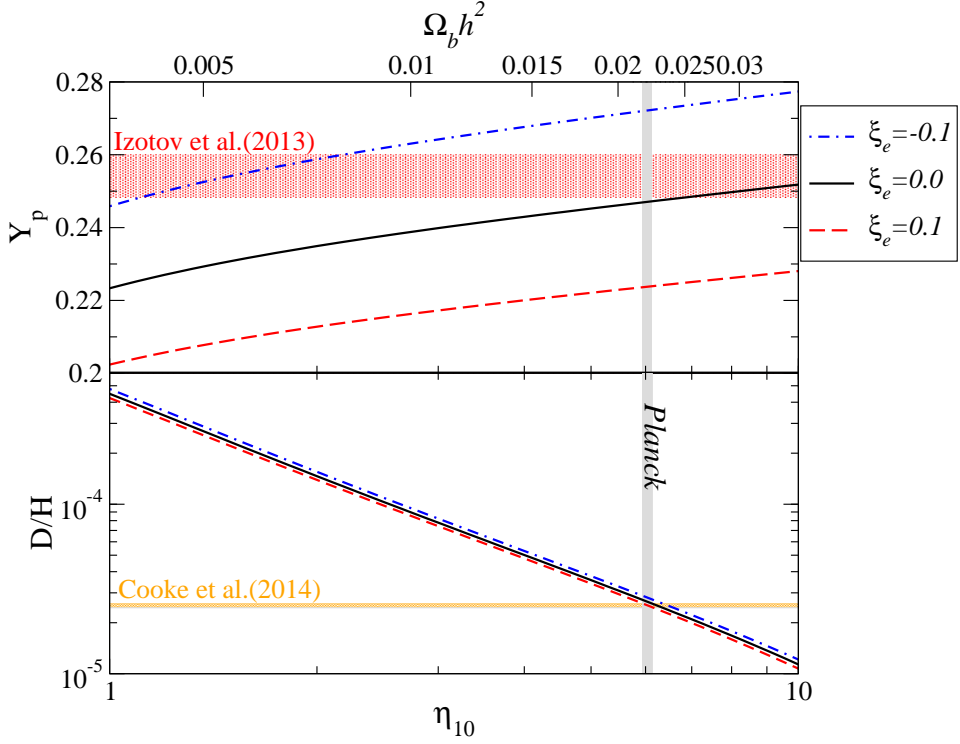}}
 \caption{Effects of neutrino degeneracy on the production of $^4$He and D/H. 
The degeneracy parameters is taken to be 
$\xi_{\rm e}=-0.1, 0$, and $0.1$ from the top to bottom
 curve. The vertical  band comes from the baryon density determined by $Planck$.
 The horizontal bands correspond to the observational abundances of
 ${}^4_{}$He and D/H with $2\sigma$ uncertainty\cite{Ichimasa2014}.}
\label{fig:bbn_xi_nacre2}
\end{figure}

\begin{figure}[th]
\resizebox{\hsize}{!}{\includegraphics{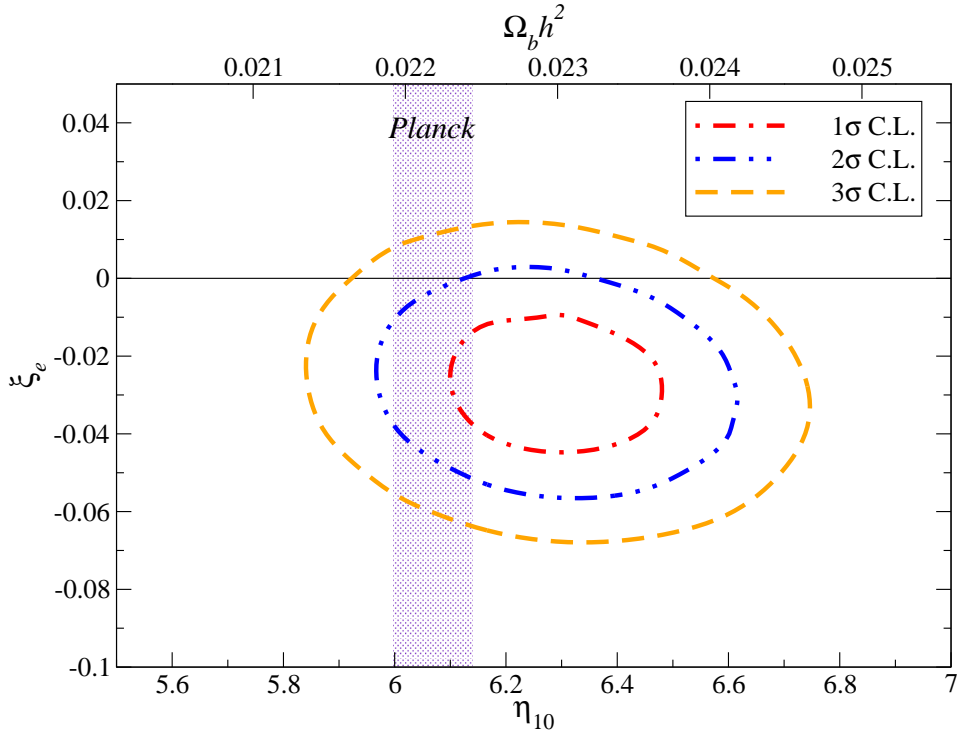}}
 \caption{Contours having 1$\sigma$, 2$\sigma$, and 3$\sigma$ confidence
 levels from $Y_p$ and D/H observations in the $\eta^{}_{10}-\xi_{\rm e}$ plane.
The horizontal line corresponds to SBBN~($\xi^{}_{e}=0$). The vertical
 band shows the baryon density from $Planck$\cite{Ichimasa2014}.}
\label{fig:cntr_xi_eta}
\end{figure}

Within the framework of general relativity, BBN can be, for example, extended to include
neutrino degeneracy (e.g. Ref.~\cite{Weinberg1972book}).
Degeneracy of electron-neutrinos is described in terms of a parameter
\begin{equation}
\xi^{}_{\rm e} = \mu^{}_{\nu,e}/k^{}_{\rm B}T^{}_\nu,
\end{equation}
where $\mu^{}_{\nu,e}$ is the chemical potential of electron  neutrinos
and $T_\nu$ is the temperature of neutrinos.
To get  abundance variations of both neutrons and protons 
caused by nonzero $\xi^{}_{\rm e}$ values,
 we take a usual method to incorporate
the degeneracy into the Fermi-Dirac distribution of neutrinos~\cite{Weinberg1972book}.
In this study, we do not consider the degeneracy of $\tau$- and $\mu$-neutrinos. 

In BBN calculations, we implemented the neutrino degeneracy as follows.
Before the temperature drops to the difference $\Delta m/k^{}_{\rm B}$
in the rest mass energies between a neutron and a
proton, they are in thermal equilibrium through the weak interaction
processes \eqref{eq:weak_reac}.
Below $k^{}_{\rm B}T = 4$~MeV, we solve the rate equations for n and p
until $k{}_{\rm B}T$ drops to 1 MeV including 
{{the individual}} weak interaction 
{{rates.}} After that,
we begin to operate the nuclear reaction network with the weak
interaction rates between n and p included.
We should note that in the present parameter range shown below, effects
of neutrino degeneracy on the expansion
and/or cooling of the universe can be almost neglected, because the
absolute values of neutrino degeneracy are
rather small, and affect the 
energy density by at most $10^{-3}\ \%$. (see Fig.\ref{fig:cntr_xi_eta}).  

The produced amounts of D and ${}^7_{}$Li are almost the same 
as those in the SBBN model, respectively, 
while
${}^4_{}$He becomes less abundant if $\xi^{}_{\rm e} > 0$, since
$\beta$-equilibrium {{leads to lower neutron production.}} 
This is because the abundance ratio of neutrons to protons (n/p) is
proportional to $\exp [- \xi_{\rm e}]$.
This can be seen in Fig.\ref{fig:bbn_xi_nacre2}; {{while}} the abundance of
 ${}^4_{}$He is very sensitive to $\xi^{}_{\rm e}$,
it is insensitive to $\eta$. On the other hand, although the abundance of D is almost uniquely
determined from $\eta$, i.e., the nucleon density, it depends weakly on
$\xi^{}_{\rm e}$. 


To find reasonable values of $\xi^{}_{e}$ and $\eta^{}_{10}$ which satisfy the consistency between
BBN and observed $^4$He and D, we
calculate $\chi^{2}_{}$ as follows:
\begin{equation}
 \chi^{2}_{}(\eta,\xi^{}_e)=\sum_{i}{\frac{\left( Y^{th}_i(\eta,\xi^{}_{e}) -
			     Y^{obs}_{i}\right)^2}
{{\sigma^{2}_{th,i}}+{\sigma^{2}_{obs,i}}}}, 
\label{eq:chisq_xi}
\end{equation}
where $Y^{}_{i}$ and $\sigma^{}_i$ are the abundances and their uncertainties
for elements $i~(i=Y_p, {\rm D})$, respectively.
The value $\sigma_{th,i}$ is obtained from the Monte-Carlo calculations using 1$\sigma$ errors associated with 
nuclear reaction rates. 
The observational values, $Y^{obs}_i$ and their errors $\sigma^{}_{obs,i}$, are taken from \eqref{eq:HeobsAver2013} and \eqref{eq:DobsPettini2012}. 

Figure~\ref{fig:cntr_xi_eta} shows the contours 
enclosing
1$\sigma$,
2$\sigma$, and 3$\sigma$ C.L. in the
$\eta^{}_{10}-\xi^{}_{\rm e}$ plane obtained from \eqref{eq:chisq_xi}.
In consequence, we get the following constraints for both $\eta_{10}$ and
$\xi_{\rm e}$\cite{Ichimasa2014}:
\begin{align}
 &6.17 < \eta^{}_{10} < 6.38 \,\,\,\ \
 -3.4\times10^{-2} <\xi_{\rm e} < -1.8\times10^{-2}~~~\text{(1$\sigma$ C.L.), }
\label{eq:xi_eta_results_1} \\
 & 6.02 < \eta^{}_{10} < 6.54 \,\,\,\
 -4.6\times10^{-2} <\xi_{\rm e} < -0.4\times10^{-2} ~~~\text{(2$\sigma$ C.L.).}
\label{eq:xi_eta_results_2}
\end{align}
It is noted that, except for neutron decay, two-body reactions are
dominant during BBN. 
Only two reaction of the
$\beta$-decay
of ${}^3$H with $\tau^{}_{1/2}=12.32$~y  
and e-capture of ${}^7$Be
with  $\tau^{}_{1/2}=53.24$~d~\cite{site:ENSDF} 
are important weak reactions of light nuclides.
These half lives are modified
by a small factor through neutrino degeneracy. However, the
final abundance is not affected at all.

\section{Inhomogeneous Big-Bang nucleosynthesis}
\subsection{Recent study of inhomogeneous BBN}
\label{sec:intro_ibbn}

 The study of SBBN has been done under the assumption of the homogeneous
 universe. On the other hand, BBN with the inhomogeneous baryon
 distribution also has been investigated. The model is called an
 inhomogeneous BBN (IBBN).  IBBN 
results from 
the inhomogeneity of baryon
 concentrations that could be induced by
 baryogenesis\cite{Matsuura:2004ss} or phase transitions such
as QCD or electro-weak phase 
transition\cite{2zone,TerasawaSato89,IBBN0,IBBN1,Jedamzik1994,Fuller1988,IBBN_QCD} 
during the expansion of the universe. 

 \begin{figure}[t]
 \begin{center}
\includegraphics[width=0.7\linewidth,keepaspectratio,clip]{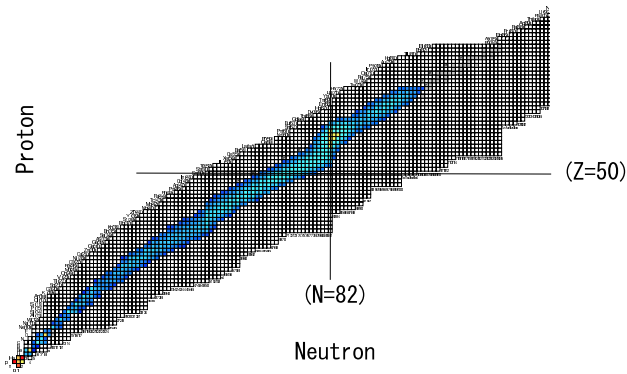}
\vspace{-0.5cm}
\caption{Abundance distribution at $T=3\times10^{9}$~K with $\eta=10^{-4}$. 
From Matsuura et al.\cite{Matsuura2007}.}
\label{fig:mo3}
\end{center}
\end{figure}

\begin{figure}[t]
\begin{center}
\includegraphics[width=0.7\linewidth,keepaspectratio,clip]{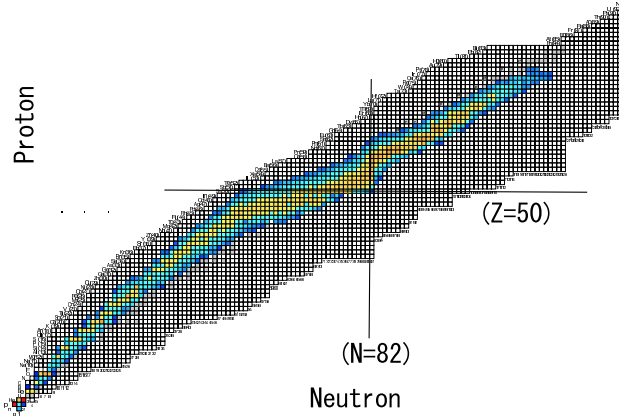}
\vspace{-0.3cm}
\caption{Abundance distribution at $T=10^{9}$~K with $\eta=10^{-4}$\cite{Matsuura2007}.}
\label{fig:mo5}
\end{center}
\end{figure}

A new astrophysical site of BBN is presented\cite{Matsuura2007} that
 contains very high $\eta$. 
Figures~\ref{fig:mo3} and \ref{fig:mo5} show 
nuclear abundance distributions in 
 a high density region with $\eta=10^{-4}$ on the nuclear chart
 covering heavy elements.
As shown in Fig.~\ref{fig:mo3}, stable nuclei  are first synthesized. 
When the temperature goes down, proton- and  neutron-capture reactions are active for the
nuclei with $A<100$ and with $A>120$, respectively (Fig.~\ref{fig:mo5}). 
Namely, it suggests that both the $r$-process and the $p$-process can occur at the BBN era. 
On the other hand, for $\eta^{}=10^{-3}_{}$, a proton capture is only active\cite{Matsuura2007}.
The problem to be solved is the origin and evolution of the high density region.
The size of the high density island 
is estimated\cite{Matsuura2007} to be $10^5 - 10^{17}$ cm at the BBN epoch.
The upper bound is obtained from the maximum angular resolution of CMB
and the lower bound is from the analysis\cite{IBBN0} of comoving diffusion length of neutrons
and protons.

Some models of baryogenesis suggest that very high baryon density
regions in the early universe.
Recent observations, however, suggest that heavy elements could
already exist in high red-shift epochs and therefore the origin of these elements
becomes a serious problem.
Motivated by these facts, we investigate BBN in very 
high baryon density regions. 
The BBN proceeds in proton rich environment, in which 
a rapid $p$-process is operative. 

However, by taking heavy nuclei into account, we find that BBN proceeds through both $p$-
and $r$-processes simultaneously. Furthermore,
$p$-nuclei such as $^{92}$Mo, $^{94}$Mo, $^{96}$Ru and $^{98}$Ru, whose origin is
not well known, are also synthesized.
The above issues should be refined and checked by investigating the possible model consistent
with available observations.


\subsection{Two-zone model of IBBN}
\label{sec:ibbn_model}

Quite interesting features\cite{Matsuura2007} have been presented for
the possibility of IBBN, but relevant parameters concerning the high and low density
regions have not yet been specified. Therefore, we explore the reasonable parameters using a
simple two-zone model.\cite{IBBN1}  
The early universe is assumed to contain high and low baryon density regions. 
For simplicity we ignore the diffusion effects.

We assume that all the produced elements are mixed homogeneously at some epoch between the end of BBN and 
the recombination era.
To construct a two-zone model, we need to define $n_{ave}, n_{high}$ and $n_{low}$ as
the average-, high-, and low-number densities of baryons, $f_v$ as the volume
fraction of the high baryon density region, and
$X^{ave}_{i}, X^{high}_i$ and $X^{low}_{i}$
as the mass fractions of element $i$ in the average-, high- and
low-density regions, respectively. 
Then the basic relations between these variables are written as 
\begin{eqnarray}
n^{}_{ave} &=& f^{}_{v}n^{}_{high}+\left( 1-f^{}_v \right)n^{}_{low},
 \label{eq:num_b} \\
 n^{}_{ave}X^{ave}_{i} &=&  f^{}_{v}n^{}_{high} X^{high}_i 
  +\left( 1-f^{}_v \right)n^{}_{low} X^{low}_{i}. \label{eq:massfrac}
\end{eqnarray}
If the baryon fluctuation is assumed to be isothermal\cite{TerasawaSato89,Alcock1987,Fuller1988},
the 
following equations are derived by dividing Eqs.~\eqref{eq:num_b} and
\eqref{eq:massfrac} by the the number density of photons $n^{}_\gamma$: 
\begin{eqnarray}
\eta^{}_{ave}  &=&
 f^{}_{v}\eta^{}_{high}+(1-f^{}_v)\eta^{}_{low},  \label{eq:eta_ave}
\\
\eta^{}_{ave}X^{ave}_{i}
 &=&
 f^{}_{v}X^{high}_{i}\eta^{}_{high}+(1-f^{}_{v})X^{low}_{i}\eta^{}_{low}.
\label{eq:Yi_ave}
\end{eqnarray}
Here three kinds of $\eta$'s are 
\[
 \eta_{ave} = \frac{n_{ave}}{n_\gamma}, \qquad 
 \eta_{high} = \frac{n_{high}}{n_\gamma}, \qquad
 \eta_{low} = \frac{n_{low}}{n_\gamma}, 
\] 
where $\eta_{ave}$ is set to be the observed value by CMB\cite{WMAP3,WMAP5,Planck2015cosmo}
: $\eta=6.1\times10^{-10}$.
Both $\eta^{}_{high}$ and $\eta^{}_{low}$ are determined from $f^{}_v$ and
the density ratio $R = n^{}_{high}/n^{}_{low}=\eta_{high}/\eta_{low}$.

We note that $\rho^{}_b$ is the average baryon density obtained
from Eq.~(\ref{eq:num_b}), and temperature $T$ is set to be homogeneous.
This assumption is critically important to build our model; otherwise we must treat
the zones to evolve separately, which involves fundamental problem as opposed to 
the cosmological principle.  

\subsection{Constraints from light element abundances}
\label{sec:ibbn_He4}

\begin{figure}[tb]
\begin{center}
 \includegraphics[width=0.72\linewidth,keepaspectratio]{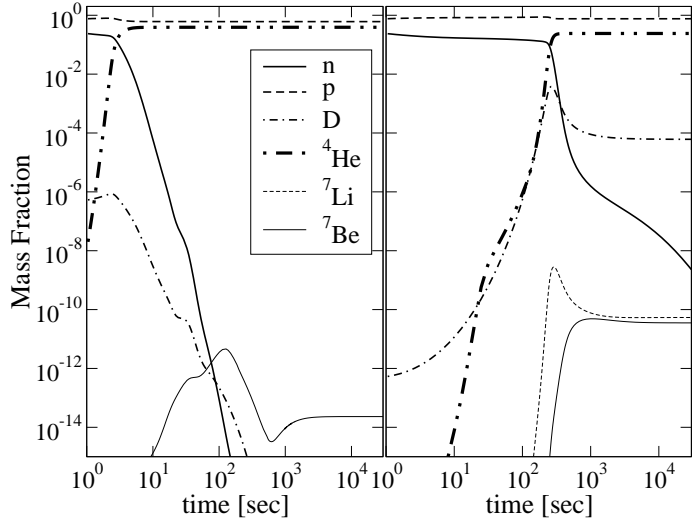}
 \caption{Evolution of light elements in IBBN with $f_v=10^{-6}$ and $R=10^{6}$. 
The left panel corresponds to 
the high density region $\eta_{high}=3.05\times10^{-4}$.
The right panel corresponds to the low density region $\eta_{low}=3.05\times 10^{-10}$.
}
\label{fig:2zone_bbn}
\end{center}
\end{figure}

We show in Fig.~\ref{fig:2zone_bbn} an example of light element synthesis 
in the high and low density regions with $f_v=10^{-6}$ and
$R=10^6$ that correspond to $\eta_{high}=3.05\times10^{-4}$ and
$\eta_{low}=3.05\times10^{-10}$.
In the right panel for 
$\eta_{low}$ the evolution of the elements is almost the
same as that of SBBN. 
In the left panel for 
$\eta_{high}$, D and $^4_{}$He are synthesized at higher
temperatures. This is because the reaction 
${\rm p} + {\rm n} \longrightarrow {\rm D} +\gamma$ starts at earlier epoch. 
In addition, the amount of ${}^4_{}$He is larger than that in
the low density region, because neutrons still remain when
the nucleosynthesis starts.
On the other hand, ${}^7_{}$Li (or ${}^7_{}$Be) is much less produced.
It implies that heavier nuclei, such as ${}^{12}$C and ${}^{16}$O, are
synthesized in the high density region.
Using these calculated abundances in both regions, we obtain the average
values of the light elements from Eq.~(\ref{eq:Yi_ave}).
Then we can put constraints on $f^{}_v$ and $R$ by comparing 
the values of $Y_{\rm p}$ and D/H with the observed abundances. 

\begin{figure}[tb]
\begin{center}
 \includegraphics[width=0.7\linewidth,keepaspectratio]{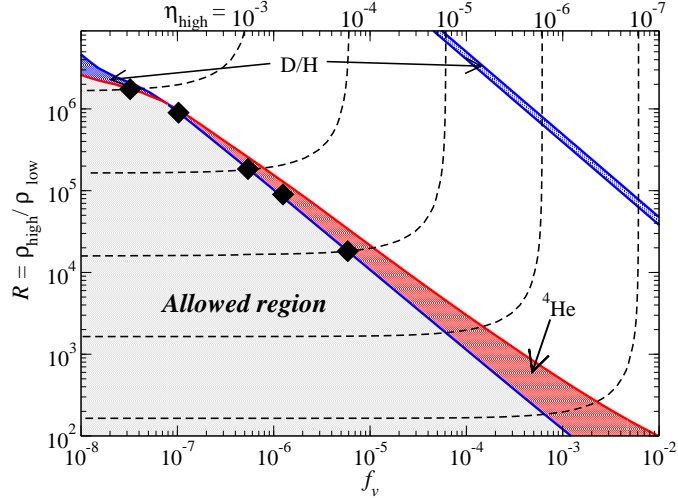}
\caption{
Constraints on the $f_v-R$ plane from the observations of ${}^4$He and D/H.
 The region below the red line is acceptable region obtained
 from ${}^4_{}$He observation by Izotov \& Thuan\cite{Izotov2010} and
 Aver et al.\cite{Aver2012}. 
 Constraints from the D/H observations\cite{Pettini2008,Pettini2012} are shown by the
 region below the blue lines. The gray region corresponds to the allowable 
parameters obtained from the two observations of ${}^4_{}$He  and D/H. 
In the narrow region bounded by two upper blue lines, only the D/H
 abundance is 
consistent with observations.
This is the contribution of the low density region with $\eta^{}_{low}\sim 10^{-12}$; 
The D abundance tends to decrease with increasing baryon density
for $\eta>10^{-12}$.  The dotted lines show 
 the contours  of the baryon-to-photon ratio in the high-density
 region. Filled squares indicate the adopted parameters in \S \ref{sec:ibbn_result}\cite{Nakamura2013}.
}
\label{fig:HeDcntr}
\end{center}
\end{figure}

In Fig.~\ref{fig:HeDcntr}, the constraints are shown in the $f^{}_v-R$
plane. Contours of calculated abundances (solid lines) and the $\eta^{}_{high}$
(dashed lines) are drawn.
In our analysis, we obtain only  the upper limit to the parameter $R$.
Note that the allowed region includes 
density values as high as $\eta_{high}=10^{-3}$. 

\begin{figure}[tb]
\begin{center}
{{
  \includegraphics[width=0.7\linewidth,keepaspectratio]{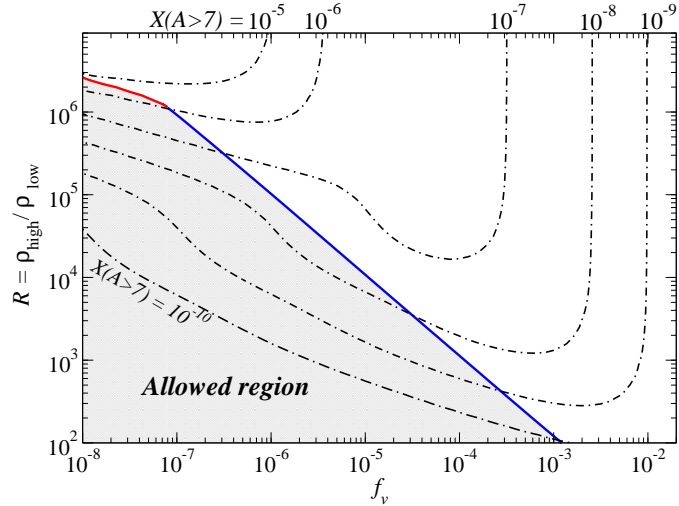}}
}
\caption{Contours of the total mass fractions of heavier nuclei ($A > 7$).
The shaded allowed region is the same as in Fig.~9. \cite{Nakamura2013}.
}
\label{fig:cntr_upLi7}
\end{center}
\end{figure}

Since $\eta_{high}$ takes a larger value, nuclei heavier
than $^7$Li are synthesized more and more. Then we estimate
the abundance of CNO elements in the allowed region.
Figure~\ref{fig:cntr_upLi7} shows the contours of the summation of 
$X^{ave}_{i}$ over heavier nuclei~($A>7$).  
As far as our small BBN code is concerned\cite{Hashimoto1985}, the total mass fraction of CNO nuclei 
amounts to $X(A>7) \simeq 10^{-5}$.

\subsection{Synthesis of heavy elements in high density region}
\label{sec:ibbn_result}

We investigate synthesis of heavy elements in the
high-density region considering the constraints shown in Fig.~\ref{fig:HeDcntr}.
The abundance change is calculated with a large nuclear reaction
network, which contains 4463 nuclei from neutron, proton to
Americium ({\it Z} = 95 and {\it A} = 292).
The nuclear data such as reaction rates, nuclear masses and partition
functions are the same as used in Ref. \citen{fujimoto} except for 
the weak interaction rates\cite{Kawano}
which is adequate for the high temperature stage $T>10^{10}$~K.
Note that the mass fractions of $^4$He and D obtained with the large
network are consistent with those  
calculated with a small network in \S\ref{sec:ibbn_He4}
within an accuracy of a few percents.

As seen in Fig.~\ref{fig:cntr_upLi7}, heavy elements are produced at the level 
$X(A>7) \ge 10^{-9}$ in the upper region of $R$ in the allowed region.
To examine the efficiency of the heavy element production, 
we select five models with parameters
$\eta^{}_{high}=10^{-3}_{}, 5.1\times10^{-4}, 10^{-4}_{}, 5.0\times10^{-5}$
, and $10^{-5}$ which correspond to
$\left( f^{}_v, R \right)= \left( 3.24\times10^{-8}, 1.74\times10^{6}_{} \right)$
, $\left( 1.03\times10^{-8}, 9.00\times10^{5}_{} \right)$
, $\left( 5.41\times10^{-7}, 1.84\times10^{5}_{} \right)$
, $\left( 1.50\times10^{-6}, 9.20\times10^{4}_{} \right)$
, and $\left( 5.87\times10^{-6}, 1.82\times10^{4}_{} \right)$, respectively.
Adopted parameters are indicated by
the filled squares in Fig.~\ref{fig:HeDcntr}.


\begin{table}[hbt]
\tbl{
Mass fractions of light elements for the two cases :
 $\eta^{}_{high}\simeq10^{-3}_{}$, $\eta_{high}=5\times10^{-4}$. 
}
{\begin{tabular}{@{}ccccccc@{}} 
\toprule
$f^{}_v, R $
& \multicolumn{3}{c}{$3.23\times10^{-8}_{}, 1.74\times10^{6}_{}$}
& \multicolumn{3}{c}{$1.03\times10^{-7}_{}, 9.00\times10^{5}_{}$}
 \\
$( \eta^{}_{high}, \eta^{}_{low} )$  
& \multicolumn{3}{c}{($1.02\times10^{-3}_{}$, $5.86\times10^{-10}_{}$) }
& \multicolumn{3}{c}{($5.10\times10^{-4}_{}$, $5.67\times10^{-10}_{}$) }
 \\
\colrule
elements & high & low & average & high & low & average \\ 
\colrule
p   & $0.586$ & $0.753$ & $0.744$ & $0.600$  &  $0.753$ &  $0.740$  \\
D   & $1.76\times10^{-21}_{}$ & $4.50\times10^{-5}_{}$ &  $4.26\times10^{-5}_{}$ 
       & $3.43\times10^{-21}_{}$ & $4.75\times10^{-5}_{}$ &  $4.34\times10^{-5}_{}$ 
\\
${}^3_{}$He+T &   $2.91\times10^{-14}_{}$ &  $2.18\times10^{-5}_{}$ & $2.07\times10^{-5}_{}$ 
	     &   $2.77\times10^{-14}_{}$ &  $2.23\times10^{-5}_{}$ & $2.04\times10^{-5}_{}$ \\
${}^4_{}$He 
& $0.413$ & $0.247$ & $0.256$
& $0.400$ & $0.247$ & $0.260$ 
\\
${}^7_{}$Li+${}^7_{}$Be &
  $1.63\times10^{-13}_{}$ &  $1.78\times10^{-9}_{}$ & $1.68\times10^{-9}_{}$ 
& $6.80\times10^{-14}_{}$ &  $1.65\times10^{-9}_{}$ & $1.52\times10^{-9}_{}$
\\ \botrule
\end{tabular}\label{tab:abundance_4463_1} }
\vspace{0.5cm}

\tbl{
Mass fractions of light elements for the two cases :
 $\eta^{}_{high}\simeq10^{-4}$, and $\eta^{}_{high}=10^{-5}$. 
}
{\begin{tabular}{@{}ccccccc@{}} 
\toprule
$f^{}_v, R $
& \multicolumn{3}{c}{$5.41\times10^{-7}_{}, 1.84\times10^{5}_{}$}
& \multicolumn{3}{c}{$5.87\times10^{-6}_{}, 1.82\times10^{4}_{}$}
 \\
$( \eta^{}_{high}, \eta^{}_{low} )$  
& \multicolumn{3}{c}{($1.04\times10^{-4}_{}$, $5.62\times10^{-10}_{}$)}
& \multicolumn{3}{c}{($1.02\times10^{-5}_{}$, $5.59\times10^{-10}_{}$)}
 \\
\colrule
elements & high & low & average & high & low & average \\ 
\colrule
p   & $0.638$  &  $0.753$ &  $0.742$ &  $0.670$ &   $0.753$ &   $0.745$ \\
D 
      & $6.84\times10^{-22}_{}$ & $4.79\times10^{-5}_{}$ &  $4.36\times10^{-5}_{}$
      & $1.12\times10^{-22}_{}$ & $4.48\times10^{-5}_{}$ &  $4.37\times10^{-5}_{}$
\\
${}^3_{}$He+T &   $1.63\times10^{-13}_{}$ &  $2.23\times10^{-5}_{}$ & $2.04\times10^{-5}_{}$ 
	     &  $1.49\times10^{-9}_{}$ &  $2.25\times10^{-5}_{}$ & $2.03\times10^{-5}_{}$ \\
${}^4_{}$He 
& $0.362$ & $0.247$ & $0.258$ 
& $0.330$ & $0.247$ & $0.254$
\\
${}^7_{}$Li+${}^7_{}$Be 
& $7.42\times10^{-13}_{}$ &  $1.64\times10^{-9}_{}$ & $1.49\times10^{-9}_{}$ 
& $6.73\times10^{-8}_{}$  &  $1.62\times10^{-9}_{}$ & $7.96\times10^{-9}_{}$
\\ \botrule
\end{tabular}\label{tab:abundance_4463_2} }
\end{table}

Tables~\ref{tab:abundance_4463_1} and \ref{tab:abundance_4463_2}
 give the abundance of light elements 
in the high and low density regions.
The mass fractions in the low density region are the
same as those obtained in \S~\ref{sec:ibbn_He4}, because the
abundance flows beyond $A=7$ are negligible.
We should note that the averages of abundances $Y_{\rm p}$ and D/H
in the two regions
coincide with the observed abundances, respectively.

\begin{figure}[tb]
\begin{minipage}{.49\linewidth}
\begin{center}
 \includegraphics[width=1.\linewidth,keepaspectratio]{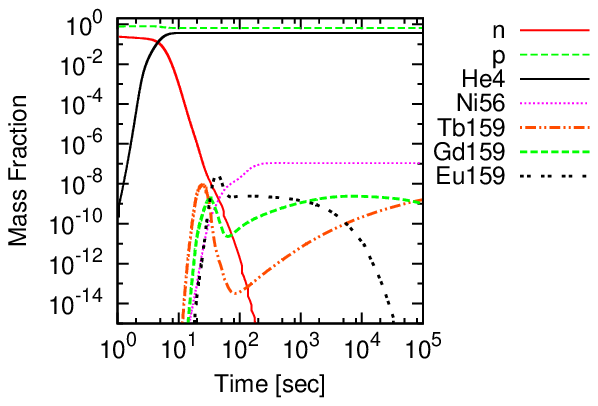}

 {(a) $\eta^{}_{high}=1.02\times10^{-4}_{}$}
\end{center}
\end{minipage}
\begin{minipage}{.49\linewidth}
\begin{center}
 \includegraphics[width=1.0\linewidth,keepaspectratio]{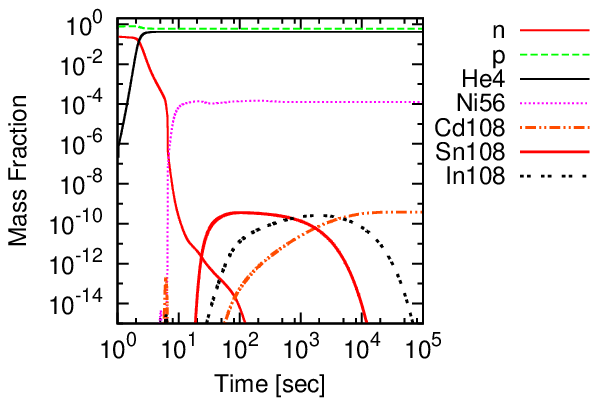}

 {(b) $\eta^{}_{high}=1.06\times10^{-3}_{}$}
\end{center}
\end{minipage}
\caption{Time evolution of the mass fractions in high-density regions of
  (a) $\eta^{}_{high}=1.02\times10^{-4}_{}$ and (b) $\eta^{}_{high}=1.06\times10^{-3}_{}$, respectively. }
\label{fig:mpure}
\end{figure}

Figure~\ref{fig:mpure}(a) shows the result of nuclsosynthesis in 
the high density region of $\eta^{}_{high}=1.04\times 10^{-4}$. The nucleosynthesis paths 
proceed along the stability line during a few seconds, and afterwards they are 
classified with the mass number.
 For nuclei with $A \leq 100$, proton captures become
 very active  compared to neutron capture at $T>2\times10^{9}$~K and
 the path shifts to the proton rich side, which begins from breaking out
 the hot CNO cycle.
 For nuclei of $100 < A < 120$,
 the path goes across the stable nuclei from proton- 
 to neutron-rich side, since temperature decreases and the number of
 seed nuclei of neutron capture increases significantly. 
 Neutron captures become much more efficient for heavier nuclei of ${\it A} \ge 120$. 
The neutron capture is not
 similar to the canonical $r-$process, since the nuclear reactions 
proceed under the condition of the high-abundance of protons. 
For example, $^{159}$Tb, $^{159}$Gd and $^{159}$Eu are synthesized 
through neutron captures. 
After $t=10^3$ sec, we can see $\beta$-decays
$^{159}$Eu $\rightarrow \,^{159}$Gd $\rightarrow \,^{159}$Tb,
where the half life of $^{159}$Eu and $^{159}$Gd are 
$18.1$~min 
 and $18.479$ h\cite{site:ENSDF}, respectively. 

 The results of $\eta^{}_{high}=1.06\times 10^{-3}_{}$ is shown in Fig.\ref{fig:mpure}~(b).
 The reactions also first proceed along the stability
line in the high density region. 
Subsequently, the reactions directly proceed to the
 proton-rich side through rapid proton captures.
 We can see $\beta$-decays 
 $^{108}$Sn $\rightarrow \, ^{108}$In $\rightarrow \, ^{108}$Cd,
 where the half life of $^{108}$Sn and $^{108}$In are 
$10.3$ min and  $58.0$ min\cite{site:ENSDF}, respectively.
 In addition, radioactive nuclei $^{56}$Ni $(\tau^{}_{1/2}=6.075$~d\cite{site:ENSDF})
and $^{57}$Co ($\tau^{}_{1/2}=271.75$~d\cite{site:ENSDF}) are produced just after 
the formation of $^4$He
in the extremely high density
region with $\eta^{}_{high}\geq 10^{-3}$ 
like the beginning of supernova explosions.~\cite{Hashimoto1995}

\begin{figure}[t]
\begin{center}
 \includegraphics[width=.70\linewidth,keepaspectratio]{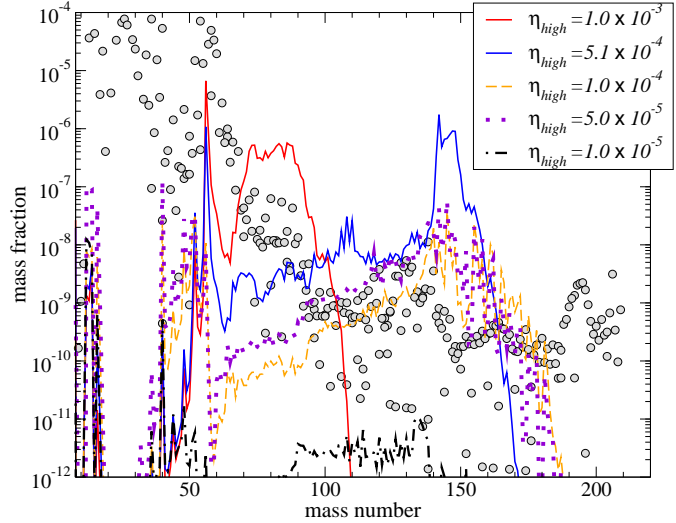}
\end{center}
\caption{Comparison of the averaged mass fractions in the two-zone model with the
 solar system abundances~\cite{Anders1989} (indicated by large dots).\cite{Nakamura2006}} 
\label{fig:massfrac_obs}
\end{figure}

Figure~\ref{fig:massfrac_obs} shows the comparison between the average mass
fraction produced in IBBN calculation and the solar system abundances by 
Anders \& Grevesse\cite{Anders1989}.
There are over-produced elements
around $A=150$ ($\eta^{}_{high}=10^{-4}$) and $A=80$ 
($\eta_{high}=10^{-3}$).
Although it seems to conflict with the chemical evolution in the universe,
this problem could be solved by the careful choice of $f_v$ and/or $R$.\cite{Nakamura2013}

{{{
In IBBN model, the lithium-7 can be synthesized in both regions depending on $f^{}_v$ and $R$ as shown in 
Tables~\ref{tab:abundance_4463_1} and \ref{tab:abundance_4463_2}. 
For $\eta^{}_{high}=1.0\times 10^{-5}$, 
the averaged value of ${}^7$Li/H is $1.5 \times 10^{-9}$, which is higher than 
the predicted value in SBBN with $\eta^{}_{10}=6.1$. 
On the other hand, in cases of $\eta^{}_{high}=5.1\times 10^{-4}$ and 
$1.0\times 10^{-4}$, the average of ${}^7$Li/H is lower than the SBBN value.
Although these ${}^7$Li/H values is still higher than the results of recent observation \eqref{eq:Li7_Sbord} , 
 there remains possibility to solve the ``Lithium problem"
  by extending our IBBN model to another kind of model such a multi-zone model.
}}}

\section{Brans-Dicke cosmology with a variable cosmological term}
\subsection{Field equation}	
The action in the Brans Dicke theory modified with a variable
cosmological term $\Lambda$ (BD$\Lambda$) which is a function of a scaler field
$\phi$, is given by Endo \& Fukui \cite{Endo} as
\begin{equation}\label{fff}
S=\int d^{4}x\sqrt -g \left[\left(R-2\Lambda \right)\phi-\frac{\omega}{\phi}\phi_{,\nu}\phi^{,\nu}+16\pi L_{m}\right],
\end{equation}
where $R$, $L_{m}$ and $\omega$ are the scalar curvature, the Lagrangian density of matter
, and the dimensionless constant of Brans-Dicke gravity, respectively. 
The field equations for $BD\Lambda$ are written as follows\cite{Arai}: 
\begin{eqnarray} \label {p13_ee}\nonumber
R_{\mu\nu}-\frac{1}{2}g_{\mu\nu}R+g_{\mu\nu}\Lambda & = &
 \frac{8\pi}{\phi}T_{\mu\nu}+\frac{\omega}{\phi^{2}}\left(
\phi_{,\mu}\phi_{,\nu} 
-\frac{1}{2}g_{\mu\nu}\phi_{,\alpha}\phi^{,\alpha}\right) \\
& & + \frac{1}{\phi}\left(\phi_{,\mu;\nu}-g_{\mu\nu}\Box\phi\right),
\end{eqnarray}
\vspace{0.2mm}
\begin{equation}\label{aa}
R-2\Lambda-2\phi\frac{\partial\Lambda}{\partial\phi}=\frac{\omega}{\phi^{2}}\phi_{,\nu}\phi^{,\nu}-\frac{2\omega}{\phi}\Box\phi,
\end{equation}
where 
$\Box$ is the d'Alembertian.
 
The expansion is described by the following equation derived from the $(0,0)$ component of Eq. (\ref{p13_ee}):
\begin{equation}\label{p14_d}
  \left(\frac{\dot{a}}{a}\right)^{2}=\frac{8\pi\rho}{3\phi}-\frac{k}{a^{2}}+\frac{\Lambda}{3}+\frac{\omega}{6}\left(\frac{{\dot{\phi}}}{{\phi}}\right)^{2}-\frac{\dot{a}}{a}\frac{{\dot{\phi}}}{{\phi}}.
\end{equation}
   
 We adopt the simplest case of the coupling between the scalar and matter field
 \begin{equation}\label{dddd}
   \Box\phi=\frac{8\pi\mu}{2\omega+3}T^{\nu}_{\nu},
 \end{equation}
where $\mu$ is a constant. 
Original Brans-Dicke theory is deduced for $\mu=1$.

Assuming a perfect fluid for $T_{\mu\nu}$,
Eq.~(\ref{dddd}) reduces to the following:
\begin{equation}\label{eeee}
    \frac{d}{dt}\left(\dot{\phi}a^{3}\right)=\frac{8\pi\mu}{2\omega+3}\left(\rho-3p\right) a^{3}.
\end{equation} 
  Then, Eq. (\ref{eeee}) is integrated to give 
\begin{equation}\label{p14_g}
 \dot{\phi}=\frac{1}{a^{3}}\left[\frac{8\pi\mu}{2\omega +3}\rho_{{m}_{0}}t+B \right],   
\end{equation}where $B$ is an integral constant and from now on we use the normalized value of $B$: $B^{*}=B/(10^{-24} {\rm{ g \ s \ cm^{-3}}})$.
	
A particular solution of Eq. (\ref{aa}) is obtained from Eqs. (\ref{p13_ee}) and (\ref{dddd}):  
\begin{equation}\label{p14_h}
 \Lambda=\frac{2\pi\left(\mu-1 \right)}{\phi}\rho_{m_{0}}a^{-3}, 
 \end{equation}
where $\rho^{}_{m_{0}}$ is the matter density at the present epoch.
 
\begin{figure}[tb]
\begin{center}
\rotatebox{-90}{
 \includegraphics[width=7.5cm,keepaspectratio]{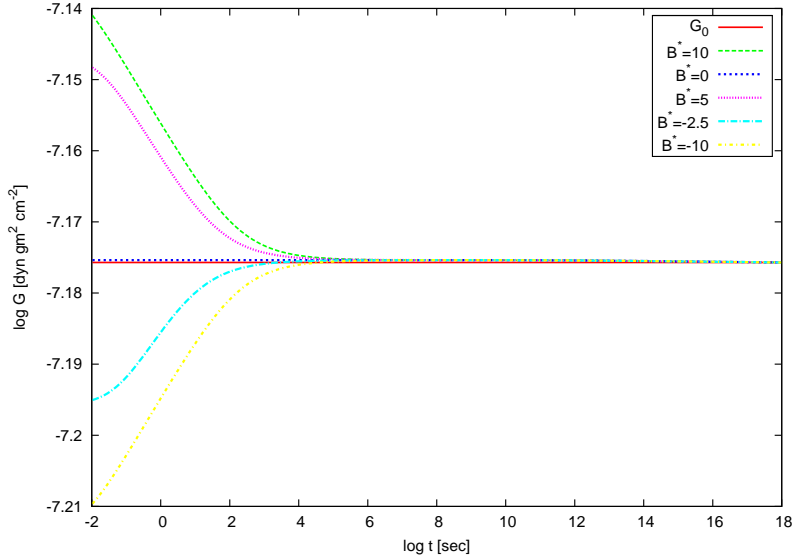}
}
\caption{Time evolution of $G$ in BD$\Lambda$ with $\mu=0.6$. $G_0$ is the
 present value:  $G_0=6.6726\times 10^{-8}~{\rm{cm^{3} g^{-1}s^{-2}}}$}
\label{fig:G-t}
\end{center}
\end{figure}

 The gravitational ``constant" $G$ is expressed as follows, 
 \begin{equation}\label{p14_hh}
  G=\frac{1}{2}\left(3-\frac{2\omega+1}{2\omega+3}\mu\right)\frac{1}{\phi}.  
 \end{equation}     
The difference of $G$ between the BD$\Lambda$  and the Friedmann model
appear at $0.01~{\rm sec} < t < 1000$~sec as shown in
Fig.~\ref{fig:G-t}.  It suggests that the expansion rate in the BBN
epoch is markedly  different.

\begin{figure}[tb]
   \centering
  \rotatebox{-90}{\includegraphics[width=7.5cm,keepaspectratio]{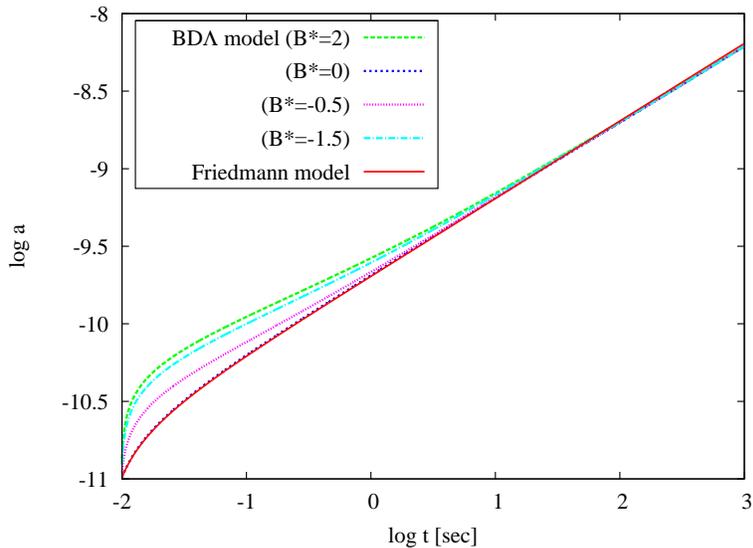}}
  \vspace{0.4cm}
     \caption{Time evolutions of the scale factor in $BD\Lambda$ with $\mu=0.6$ which are compared to the Friedmann model. }
        \label{fig:a-t}
   \end{figure}

It is reduced to
the Friedmann model when $\phi$ = constant and $\omega \gg 1$.
Physical parameters have been used to solve Eqs.~(\ref{p14_d}),
(\ref{p14_g}), and (\ref{p14_h}): 
$G_{0} = 6.6726\times 10^{-8} {\rm{cm^{3} g^{-1}s^{-2}}}$, and
$H_{0}=71~{\rm{km \ s^{-1} \  Mpc^{-1}}}$. 

The parameter $\omega$ is an intrinsic parameter in the Brans-Dicke gravitational model, and 
it determines the expansion rate of the early universe.  
In our study, we set $\omega=10000$.  Although the variations are limited to the earlier ara, still larger value of $\omega$
can be permitted for our BD$\Lambda$ model (see Fig.~\ref{fig:G-t}).
Recent observations by the Cassini measurements of the
Shapiro time delay suggested that the lower limit of $\omega$ is very
large: $\omega\geq4\times10^{4}$ (Berti et al.\cite{Berti2003}, Bertotti et al.\cite{Bertotti2005}). 

Figure \ref{fig:a-t} shows the evolution of the scale
factor in $BD\Lambda$ for the several values of $B^{*}$. We identify
considerable deviations in $BD\Lambda$ from the Friedmann model at
$t<100$ s, which depends on the specific parameters.
We note that the evolution of the scale factor depends on both the
initial value of $\phi$ and a parameter $\mu$. This is the reason why
the scale factor evolves rather differently compared to the Friedmann model.

\subsection{Parameter constraints from Big Bang nucleosynthesis}


\begin{figure}[tb]
\vspace{0.9cm}
\centering
\rotatebox{-360}{\includegraphics[width=8.7cm,height=9.6cm]{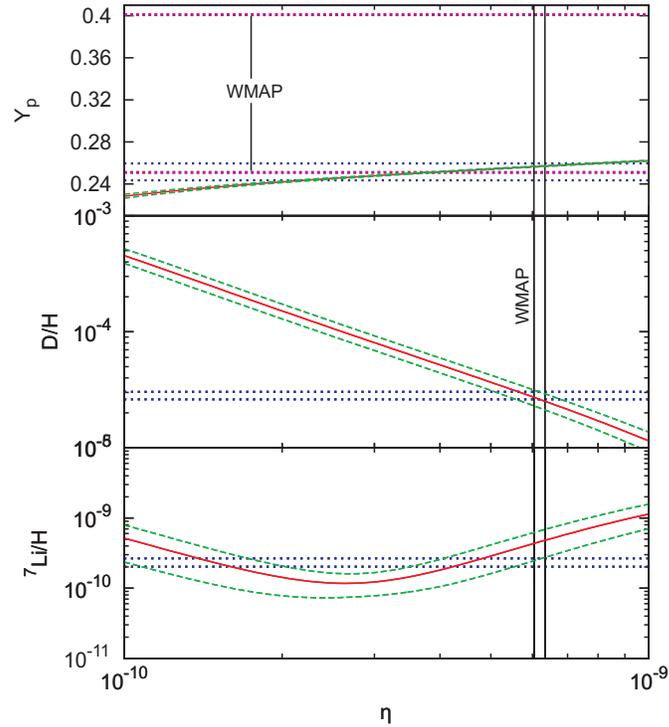}} 
\vspace{0.6cm}
\caption{Light element abundances of $\rm{{}^{4}He}$, $\rm{D}$, and $\rm{{}^{7}Li}$ vs. $\eta$ for $B^{*}=2$, $\mu=0.6$, and $\omega=10^{4}$. Dashed lines indicate the $\pm 2\sigma$  uncertainties in nuclear reaction rates in each abundance. The horizontal dotted lines indicate the regions of observational abundances. The solid vertical lines indicate the baryon-to photon ratio $\eta$.} 
\label{bbn}
\end{figure}
Big-Bang nucleosyntheis provides powerful constraints on possible
deviation from the standard cosmology\cite{Malaney}. 
Changes in the expansion as shown in Fig.~\ref{fig:a-t} affects the
abundances of light elements, because the n/p ratio
is sensitive to the expansion rate  in the BBN epoch.

  Figure \ref{bbn} shows the calculated abundances of $\rm{{}^{4}He}$, $\rm{D}$, and $\rm{{}^{7}Li}$ 
for $B^{*}=2$ and $\mu=0.6$.
The $\pm 2\sigma$ uncertainties in nuclear reaction rates
are indicated by the dashed lines. The horizontal dotted lines indicate
the observational values of $\rm{{}^{4}He}$, $\rm{D/H}$, and
$\rm{{}^{7}Li/H}$ as follows:
${Y_{\rm p} = 0.2516 \pm 0.0080}$\cite{Fukugita}, 
${Y_{\rm p}= 0.326\pm0.075}$\cite{WMAP5},
$\rm{D/H = (2.82\pm0.21)\times10^{-5}}$ (Pettini et al.\cite{Pettini2008}),
$\rm{{}^{7}Li/H=(2.34\pm0.32)\times10^{-10}}$ (Melendez \& Ramirez. \cite{Melendez}). Here two observational values of $\rm{{}^{4}He}$ are used. The solid vertical lines indicates the WMAP constraint of the baryon-to-photon ratio,  $\rm{\eta=(6.19\pm0.15)\times10^{-10}}$ (Komatsu et al.\cite{WMAP5}).

 The intersection range of the two observational values of
 $\rm{{}^{4}He}$ is used to constrain the parameters.   
It is found that the values of $\eta$ derived from $\rm{{}^{4}He}$ and
 $\rm{D/H}$ are tightly consistent with 
the value by WMAP, 
while that from ${}^7_{}$Li/H is barely consistent. We have a very small
 parameter region for ${}^7_{}$Li/H where the bands of the theoretical
 abundance, the observational abundance, and the WMAP eta overlap.
These agreements lead us to 
the parameter ranges 
of $0.0\leq~\mu\leq0.6$ and $-2\leq B^{*}\leq2$.

\section{Summary and discussions}
\label{rec:summary}

We  have reviewed the SBBN, 
BBN with the neutrino degeneracy,
IBBN, and 
BBN in the Brans-Dicke cosmology.
 First, we have compared the results of SBBN with the current observations of light elements. 
Considering the uncertainties in the nuclear reaction rates and the observed errors, 
we can summarize as follows: 
\begin{enumerate}
\item  The consistency is confirmed as far as ${}^4$He and D are concerned.
\item Large uncertainties of He observations still allow the possibility that some unknown processes beyond SBBN affected the primordial abundance.
 For example, a large value of $Y_{\rm p} \simeq 0.3$ has been reported in
low metallicity stars in globular clusters~\cite{Izotov2010}.  This large amount of He could be ascribed to the local neutrino degeneracy.
\item The significant discrepancy concerning $^7$Li remains to be solved. 
\end{enumerate}

Second, consistency between IBBN and the observations of ${}^4$He and D/H abundances has been investigated
under the thermal evolution of the standard model with $\eta_{_{\rm WMAP}}$.
We have examined the two-zone model, where the universe has the high and
low baryon density regions separately at the BBN epoch. 
We have calculated nucleosynthesis that covers 4463 nuclei in the high density region.
Below we summarize our results 
and give some prospects.
\begin{enumerate}
\item There are significant differences of 
the evolution of the
light elements between the high and low density regions;
In the high density region, nucleosynthesis begins at higher temperature.
$^4$He is more abundant than that in the low density region. 
\item Both $p$- and $r$-elements are synthesized
simultaneously in the high density region with $\eta_{high}\simeq 10^{-4}$.
Total mass fractions of 
nuclides heavier
than $^7$Li amount to 
$10^{-7}$ for $\eta_{high}=10^{-4}$ and $10^{-5}$ for $\eta_{high}=10^{-3}$.
The average mass fractions in IBBN are comparable to the solar system abundances.
\item Heavy elements beyond Fe surely affects the formation process of 
the first generation stars due to the change in the opacity.
Therefore, it may be also necessary for IBBN to be constrained from
the star formation scenarios.

\item 
{{The observed abundances of $^7$Li~\cite{sbord} cannot be explained in terms of SBBN.
Although our IBBN model is also inconsistent with $^7$Li observations, the theoretical value of  $^7$Li
is lower than that  of SBBN. Modifications of IBBN model may conduce to the answer. }}
\end{enumerate}
\
Third, we show a possible alternate theory of gravity, Brans-Dicke
theory with a variable $\Lambda$ term. Even for a very large coupling
constant $\omega$, this theory can 
explain the observed primordial abundances of the light elements.
Therefore, it is still
worthwhile to continue to explore some non-standard cosmology.

\section*{Acknowledgments}

This work was supported by JSPS　KAKENHI  Grant Numbers　JP24540278 and JP15K05083.

\bibliographystyle{ws-mpla}


\end{document}